\documentclass[conference]{IEEEtran}

\usepackage{amsmath,amssymb,amsfonts}
\usepackage{graphicx}
\usepackage{textcomp}
\usepackage{booktabs}

\usepackage[bookmarks=true,hidelinks]{hyperref}

\hypersetup{
  pdftitle={Per-Platform GPIO Overhead in Hardware-Validated Edge ML Inference Timing},
  pdfauthor={Akul Swami, Nikhil Chougule},
  pdfsubject={Edge ML inference timing characterization},
  pdfkeywords={edge machine learning, ECG classification, timing validation, embedded systems, GPIO instrumentation}
}

\begin{document}

\title{Per-Platform GPIO Overhead in Hardware-Validated
Edge ML Inference Timing
\\
\large{\textit{Work in Progress}}}

\author{
\IEEEauthorblockN{Akul Swami}
\IEEEauthorblockA{Independent Researcher\\
San Jose, CA, USA\\
swami.akul@alumni.uml.edu}
\and
\IEEEauthorblockN{Nikhil Chougule}
\IEEEauthorblockA{Independent Researcher\\
Reno, NV, USA\\
chougulenikhiljgd@gmail.com}
}

\maketitle

% ---------------------------------------------------------------------------
\begin{abstract}
Edge machine learning (ML) deployments increasingly rely on
per-inference timing measured by software clocks such as Python's
\texttt{perf\_counter}, but these measurements are not always
validated against external hardware references on embedded
Linux, and edge ML benchmarking methodologies typically do not isolate
platform-dependent instrumentation overhead. This paper reports a preliminary characterization of GPIO call
overhead in hardware-validated edge
ML inference timing on two embedded platforms running a one-dimensional convolutional neural network
(1-D CNN) arrhythmia classifier on electrocardiogram (ECG) data
from the MIT-BIH Arrhythmia Database, with five classes per the
Association for the Advancement of Medical Instrumentation
(AAMI) EC57 standard. Across
$n = 10$ trials on each platform at a controlled
steady-state baseline, the per-platform constant on the Jetson Orin Nano
(TensorRT FP16, \texttt{Jetson.GPIO}) is approximately
$-20\,\mu$s, and on the Raspberry Pi 4
(ONNX Runtime CPU, \texttt{pigpio}) approximately
$-86\,\mu$s, yielding a cross-platform asymmetry of
approximately $66\,\mu$s that is large relative to
commonly used uniform validation tolerances. The Jetson constant is
well-approximated by direct GPIO call duration (the direct profile
accounts for ~88\% of the platform constant), while the Pi direct
profile over-predicts the platform constant by ~19\%, motivating
empirical per-platform calibration in the deployed measurement
context. The Pi constant is not a single sharp value but exhibits a
cross-day range of approximately $6\,\mu$s across the three
sessions sampled, while the Jetson constant reproduces to
within approximately $0.14\,\mu$s. These preliminary results suggest that cross-platform edge ML
timing studies may benefit from platform-aware and potentially
session-aware validation gates.
\end{abstract}


% ---------------------------------------------------------------------------
\section{Introduction}
\label{sec:introduction}

Edge machine learning systems increasingly run on commodity embedded
Linux platforms such as the NVIDIA Jetson family and the Raspberry Pi,
where per-inference timing is collected with software clocks like
Python's \texttt{perf\_counter}, POSIX \texttt{clock\_gettime}, or
equivalent monotonic timers. These measurements support inference
benchmarking, real-time deployment validation, and cross-platform
comparison. Edge ML benchmarking methodologies such as MLPerf
Inference~\cite{mlperf_inference} report latency at millisecond and
sub-millisecond granularity but treat the software-clock
measurement as ground truth. Related work on ANN inference response time on embedded
systems addresses clock-rate and threading
effects~\cite{huber_response_time_2025} but does not isolate
GPIO call overhead.

Two assumptions deserve scrutiny. First, per-inference timing is
rarely validated against an external hardware reference on the
deployed platform, leaving the systematic bias between software-clock
intervals and on-wire signal timing uncharacterized. This bias
is largely attributable to GPIO instrumentation call overhead
rather than software-clock inaccuracy, and its magnitude is
platform-dependent at a level large relative to typical uniform
validation tolerances. Second, when
measurements are compared across platforms, a uniform tolerance band
is implicitly assumed; this assumption appears not to hold for the two edge ML
platforms examined here.

This work targets cross-platform benchmarking methodology, where
logic-analyzer instrumentation is feasible as a calibration step;
in-production deployment validation without external hardware
reference is out of scope. This work reports a preliminary
characterization
of GPIO call overhead in hardware-validated inference timing on Jetson
Orin Nano and Raspberry Pi~4, using a 1D CNN ECG arrhythmia
classifier trained on the MIT-BIH Arrhythmia
Database~\cite{mitbih_physionet,physionet_goldberger} as
representative workload. The reference instrument is a Saleae Logic Pro 8 USB logic
analyzer sampling the GPIO pin at $100$\,MHz, providing
sub-microsecond edge resolution independent of the embedded
host's software clocks. The per-platform constants measured
differ by approximately $66\,\mu$s; this work characterizes their
sensitivity to a post-process \emph{glitch filter} (a
configurable dwell-time threshold below which spurious edges are
discarded before pulse pairing) and compares them to direct GPIO
call-duration profiling. The Jetson constant is well-explained by
direct call duration; the Pi constant is not, motivating empirical
calibration in the deployed context. Cross-platform edge ML timing
studies require both platform-aware and session-aware validation gates.

% ---------------------------------------------------------------------------
\section{System and Methodology}
\label{sec:methodology}

\subsection{Platforms}
\label{sec:platforms}

We characterize per-inference timing on two embedded Linux platforms
representative of current edge ML deployment targets. The Jetson Orin
Nano Super (8\,GB) runs JetPack~6.2.2 in 25\,W \emph{Super Mode}, with
the CPU governor pinned to performance (1.728\,GHz on all six cores) and
\texttt{jetson\_clocks} engaged to pin the GPU at 1.020\,GHz. Inference
on Jetson uses a TensorRT FP16 engine. The Raspberry Pi~4 Model~B
(4\,GB) runs Raspberry Pi OS Lite 64-bit (Debian~13), with the CPU
governor pinned to performance (1.8\,GHz on all four cores) and the
\texttt{pigpio} daemon (v79, built from source) handling GPIO control.
Inference on Pi uses ONNX Runtime CPU. Two cross-platform differences
are deliberate: the GPIO driver (\texttt{Jetson.GPIO} chardev under
\texttt{sudo} on Jetson; \texttt{pigpio} client over a Unix domain
socket to a daemon on Pi) and the inference backend. The inference
backend is platform-appropriate and does not interact with GPIO call
overhead, which is instead determined by the GPIO call path bracketing the
inference call.
Section~\ref{sec:direct-gpio} confirms the per-platform
asymmetry empirically in the absence of an inference workload.

\subsection{Workload}
\label{sec:workload}

The classifier is a 1-D CNN with approximately $115$\,k
parameters trained on the MIT-BIH Arrhythmia
Database~\cite{mitbih_physionet,physionet_goldberger} for
arrhythmia classification under the AAMI EC57 standard.
EC57 maps raw MIT-BIH beats to five clinically meaningful
classes (normal, supraventricular ectopic, ventricular ectopic,
fusion, unknown). The patient-disjoint training (DS1) and
evaluation (DS2) split of de Chazal et al.~\cite{dechazal2004}
is adopted, in which the 48
MIT-BIH records are partitioned into two halves of 22 records
each (excluding paced records) such that no patient appears in
both sets. Each trial replays the same 407-beat
DS2 subset in the same order, producing 407 inferences per trial and
one GPIO pulse pair per inference.

\subsection{Per-Inference Pulse Protocol}
\label{sec:pulse-protocol}

The geometry of the timing measurement is load-bearing for the rest of
the paper. Around each inference call, the orchestrator captures four
monotonic timestamps in this order:

\begin{verbatim}
t0 = perf_counter_ns(); gpio.high(PIN)
t1 = perf_counter_ns(); y = model.infer(x)
t2 = perf_counter_ns(); gpio.low(PIN)
t3 = perf_counter_ns()
\end{verbatim}

The orchestrator records $\texttt{latency\_perf\_ns} = t_2 - t_1$ as
the per-inference \texttt{perf\_counter} latency. A Saleae Logic Pro~8
captures the GPIO pin at 100\,MHz; rising and falling edges are paired
in strict order during post-process to yield
$\texttt{latency\_saleae\_ns}$ per inference.

Critically, the rising edge on the wire occurs \emph{after}
\texttt{gpio.high} is called but \emph{before} $t_1$ is captured;
symmetrically, the falling edge occurs \emph{after} $t_2$ but
\emph{before} \texttt{gpio.low} returns. The wire-edge interval is
\emph{strictly contained inside} the perf\_counter interval.

Define the per-inference timing residual
\begin{equation}
\Delta = \texttt{latency\_saleae\_ns} - \texttt{latency\_perf\_ns}
\label{eq:delta}
\end{equation}
as the signed difference between the wire-edge and software-clock
intervals for a given inference. Under the geometric containment
described above, $\Delta$ is expected to be nonpositive,
modulo scheduler preemption between \texttt{gpio.high} and the
subsequent \texttt{perf\_counter\_ns} on a non-real-time kernel.
Across the 2{,}030 inferences in the calibrated-state dataset
(5 trials at 406 post-warmup inferences each), zero positive
$\Delta$ values are observed; the cross-trial maximum is
$-18.90\,\mu$s, consistent with empirical containment. The magnitude
$|\Delta|$ is bounded above by the sum of the two GPIO call durations
plus the small internal delays between call return and wire edge.

\subsection{Operating State and Validation Conditions}
\label{sec:operating-state}

The per-platform constants reported here are stable only under a
specific operating-state envelope re-asserted each session. On the Jetson, this comprises Super Mode
(\texttt{nvpmodel -m 1}), the CPU governor pinned to
performance, and \texttt{jetson\_clocks} engaged. On the Pi,
it comprises the CPU governor pinned to performance and an
active \texttt{pigpiod} daemon. This is referred to collectively as the ``calibrated state.'' Each trial is preceded by an inter-trial
thermal cooldown to a constant entry temperature ($N_{\mathrm{jetson}}
= 29\,^\circ$C, $N_{\mathrm{pi}} = 12\,^\circ$C above ambient,
exceeding measured idle steady-state delta by $\sim 3\,^\circ$C),
ensuring constancy of thermal entry state across trials. All
measurements are at C0 baseline (no contention workload); behavior
under contention is left to future work.

\subsection{Calibration and Validation}
\label{sec:methodology-validation}

The per-platform constant $C_p$ is treated as the empirically
calibrated
central tendency of $\Delta$ on platform $p$ under calibrated state at
C0:
\begin{equation}
C_p = \operatorname{mean}_{i \in \mathcal{T}_p} \left( \operatorname*{median}_{j \in \mathcal{I}_i} \Delta_{i,j} \right)
\label{eq:cp}
\end{equation}
\noindent where $\mathcal{T}_p$ is the trial set on platform $p$ and
$\mathcal{I}_i$ the inference set within trial $i$. The inner median
provides robustness to per-inference outliers (e.g.\ infrequent
preemption events); the outer mean preserves cross-trial information.
A \emph{platform-aware} validation gate accepts a measurement chain on
platform $p$ if $|\Delta - C_p|$ remains within a uniform residual
tolerance $\tau$. Section~\ref{sec:implication} argues that no single
$\tau$ can be applied to $\Delta$ directly across platforms; the
per-platform subtraction is necessary.

% ---------------------------------------------------------------------------
\section{Empirical Findings}
\label{sec:findings}

Three measurements are reported: the per-platform constant $C_p$
($n=10$ per platform, one session this work) under calibrated state at C0; sensitivity
of $C_p$ to the post-process glitch filter; and direct profiling of
GPIO call duration on each platform.

\subsection{Per-Platform Constants}
\label{sec:per-platform}

Table~\ref{tab:per-platform} summarizes per-trial $\Delta$
statistics on each platform ($n=10$ per platform). The per-platform constant for the Jetson is denoted
as $C_{\mathrm{jetson}}$ and for the Pi as $C_{\mathrm{pi}}$,
each computed by Eq.~(\ref{eq:cp}) over its respective trial
set. Cell-level
operating state and trial provenance are recorded with each trial.

\begin{table}[t]
\caption{Summary of $\Delta$ statistics per platform at C0 baseline,
$n=10$ trials per platform (one session this work). Jetson runs
TensorRT FP16 with Jetson.GPIO chardev; Pi runs ONNX Runtime CPU
with pigpio. $C_p$ is the mean of per-trial medians of $\Delta$
(Eq.~\ref{eq:cp}); ``trial medians'' is the (min, max) of the
per-trial median of $\Delta$ across the 10 trials; ``std($\Delta$)''
is the (min, max) of the per-trial standard deviation of $\Delta$
across 407 inferences. All values in $\mu$s.
$|C_{\mathrm{jetson}}-C_{\mathrm{pi}}| \approx 66\,\mu\mathrm{s}$.}
\label{tab:per-platform}
\centering
\begin{tabular}{lrrr}
\toprule
Platform & $C_p$ & Trial medians & std($\Delta$) \\
\midrule
Jetson Orin Nano & $-20.00$ & $-20.41,\, -19.83$ & $2.46,\, 5.12$ \\
Raspberry Pi 4   & $-86.13$ & $-91.50,\, -82.53$ & $6.02,\, 14.79$ \\
\bottomrule
\end{tabular}
\end{table}

The platforms differ in $|C_p|$ by approximately $66\,\mu$s (Pi
$|C_p|$ is roughly $4\times$ larger), an order of magnitude beyond
within-trial std on either platform. Within-trial noise is therefore
unlikely to explain the asymmetry observed here: the measurement
chains appear to be systematically biased by different amounts.

Within-session trial-to-trial variation is small on Jetson (the ten
trial medians span $0.58\,\mu$s) but larger on Pi (medians span
$8.97\,\mu$s, comparable to within-trial std). Cross-day measurements
pool today's session with prior calibrated-state captures: Jetson's
two-session reproduction agrees to within approximately $0.14\,\mu$s
(prior-session median-of-medians $-20.14\,\mu$s at $n=2$, this session
$-20.00\,\mu$s at $n=10$). Pi's three sessions yielded session-level
$C_{\mathrm{pi}}$ values of $-92.12\,\mu$s ($n=2$, earliest),
$-86.73\,\mu$s ($n=3$, intermediate), and $-86.13\,\mu$s ($n=10$,
this work), spanning approximately $6\,\mu$s across the 25 trial
medians sampled. The two most recent Pi sessions agree to within
approximately $0.6\,\mu$s; the earliest session sits approximately
$6\,\mu$s lower. Whether this gap reflects genuine session-to-session
drift or undocumented protocol variance in the earliest session cannot
be distinguished with current data. Pi's per-platform constant is
therefore an empirical band rather than a single sharp value, in
contrast to $C_{\mathrm{jetson}}$ which admits a single-value
treatment; the implication is discussed in
Section~\ref{sec:implication}.

\subsection{Sensitivity to Glitch Filter Threshold}
\label{sec:glitch-sensitivity}

Saleae post-process discards filtered edges shorter than a
configurable dwell threshold (the ``glitch filter'') before
strict-ordering pulse pairing. Choice of threshold is a methodological
knob; sensitivity is characterized empirically in
Table~\ref{tab:glitch}, sweeping 13 filter values from $0$ to
$2000$\,ns on the Table~\ref{tab:per-platform} captures;
representative values are listed.

\begin{table}[t]
\caption{Glitch filter sensitivity: pulse pair counts across post-process
filter thresholds. Applied to the
Table~\ref{tab:per-platform} captures; expected pulse-pair count is
$10 \times 407 = 4070$ on each platform. Of 13 filter values swept
(0, 25, 50, 75, 100, 125, 150, 175, 200, 250, 500, 1000, 2000\,ns),
representative values are shown. ``FAIL'' indicates strict-ordering
alignment failure due to spurious edges below the filter threshold.}
\label{tab:glitch}
\centering
\begin{tabular}{rrr}
\toprule
Filter (ns) & Jetson ($n=10$) pairs & Pi ($n=10$) pairs \\
\midrule
0    & 4072 (FAIL) & 4070 \\
25   & 4071 (FAIL) & 4070 \\
75   & 4070        & 4070 \\
100  & 4070        & 4070 \\
200  & 4070        & 4070 \\
500  & 4070        & 4070 \\
2000 & 4070        & 4070 \\
\bottomrule
\end{tabular}
\end{table}

Today's Jetson capture exhibits six sub-75\,ns spurious edges
that prevent strict-ordering alignment at filter $\leq 50$\,ns; any
filter $\geq 75$\,ns suffices. Today's Pi capture has no detectable
glitches at any threshold. Spurious-edge incidence varies between
sessions (the Apr~26 captures characterized in earlier work showed
the inverse pattern, with Pi exhibiting one sub-25\,ns glitch and
Jetson none); the methodological recommendation of filter $= 100$\,ns
provides headroom on both platforms across all sessions characterized.
Within the robust band, the filter threshold varies by more than an
order of magnitude on either platform without changing median-$\Delta$
to two decimal places: per-platform constants are not sensitive to
the glitch-filter setting in these captures.

\subsection{Direct GPIO Call-Duration Profile}
\label{sec:direct-gpio}

Is $C_p$ explained by GPIO call duration alone? To test this,
a separate harness on each platform times \texttt{gpio.high} and
\texttt{gpio.low} in tight loops bracketed by
\texttt{perf\_counter\_ns} timestamps. The harness runs $n=5000$
iterations under calibrated state, with a 0.5\,ms
inter-iteration sleep, a 20-iteration warm-up, and no Saleae
capture or inference workload. The 0.5\,ms sleep is deliberately shorter than the
$\sim 1.2$\,ms inference gap of the measurement context: the harness
characterizes isolated tight-loop call duration as a comparison
baseline, not a direct simulation. Table~\ref{tab:direct-gpio} reports
the call-duration distributions.

\begin{table}[t]
\caption{Direct GPIO call-duration profile under calibrated state,
$n=5000$ iterations per platform. \textit{med(H)} and \textit{med(L)}
are median \texttt{gpio.high} and \texttt{gpio.low} call durations;
\textit{med(H+L)} is their per-iteration sum and is the closest
harness analog to $|C_p|$. All values in $\mu$s.}
\label{tab:direct-gpio}
\centering
\begin{tabular}{lrrrr}
\toprule
Platform & med(H) & med(L) & med(H+L) & $|C_p|$ \\
\midrule
Jetson & 9.83  & 8.06  & 17.89  & 20.00 \\
Pi     & 51.69 & 50.67 & 102.37 & 86.13 \\
\bottomrule
\end{tabular}
\end{table}

The two platforms tell different stories. On Jetson, HIGH+LOW
($17.89\,\mu$s) accounts for $\approx 88\%$ of $|C_{\mathrm{jetson}}|$;
the $2.11\,\mu$s residual is small and within the
uncertainty budget of the measurement chain, plausibly comprising
edge-detection latency in the logic analyzer, $10$\,ns
quantization at the $100$\,MHz sample rate, and small
orchestrator overhead between the call return and the
bracketing \texttt{perf\_counter\_ns}; the components are not
isolated here.
$C_{\mathrm{jetson}}$ is therefore mechanistically explained by call
duration plus measurement-chain overhead.

Pi is qualitatively different: the direct profile \emph{exceeds}
$|C_{\mathrm{pi}}|$ by $16.24\,\mu$s; a naive prediction of
$C_{\mathrm{pi}}$ from tight-loop profiling would be $19\%$ too high.
A verified mechanism has not been established. One plausible
hypothesis (reported here as observation, not verified claim)
is that the \texttt{pigpio} library's\footnote{\texttt{pigpio}
library documentation: \url{https://abyz.me.uk/rpi/pigpio/}.}
daemon-mediated calls exhibit different request-queue depth in
tight-loop versus inference-interleaved patterns: the orchestrator's
$\sim 1.2$\,ms inter-call gap may drain the pigpiod queue between
calls, while the harness's tight loop may allow queued requests to
overlap. Confirming this requires \texttt{pigpiod}-internal instrumentation
that has not been performed in this work. The contribution is precisely that
$C_{\mathrm{pi}}$ \emph{cannot} be predicted from profile-only
measurement and must be empirically calibrated under the deployed
measurement context.

% ---------------------------------------------------------------------------
\section{Methodological Implication}
\label{sec:implication}

The findings suggest limitations in the simplest cross-platform validation
gate: a
uniform tolerance $\tau$ applied directly to $\Delta$. Tight enough
to constrain Jetson ($|\Delta| \approx 20\,\mu$s), it rejects every
Pi measurement ($|\Delta| \approx 87\,\mu$s) as inaccurate, even
though \texttt{perf\_counter} on Pi tracks inference work as
faithfully as on Jetson. Loosened to admit Pi (e.g.\ $\tau = 100\,
\mu$s), it becomes non-binding on Jetson. The $\approx 66\,\mu$s
asymmetry makes a single $\tau$ difficult to choose if the goal
is to meaningfully constrain both platforms.

The platform-aware alternative is direct:
\begin{equation}
\text{Validation gate: } |\Delta - C_p| \leq \tau,
\label{eq:gate}
\end{equation}
where $C_p$ is the empirically calibrated per-platform constant of
Eq.~(\ref{eq:cp}) and $\tau$ is a uniform tolerance applied to the
\emph{residual}. Today's residual within-trial std is
$2.46$--$5.12\,\mu$s on Jetson and $6.02$--$14.79\,\mu$s on Pi
(Table~\ref{tab:per-platform}). Setting $\tau$ at approximately $2.5\times$ the worst
within-trial standard deviation ($14.79\,\mu$s on Pi) gives
$\tau \approx 37\,\mu$s. This threshold accepts the empirical
residual on both platforms under calibrated state, while
remaining tight enough to flag genuine measurement faults: a
fault sufficient to invalidate the timing chain (a stuck-high
GPIO line, a daemon timeout, or a kernel-scheduling pathology)
would manifest at residuals an order of magnitude beyond $\tau$.

Two empirical findings argue that $C_p$ must be obtained empirically
in the deployed context, not predicted from first principles. First,
direct profiling on Pi exceeds $|C_{\mathrm{pi}}|$ by $16.24\,\mu$s
($\approx 19\%$ over-prediction). Second, Pi's cross-day range
($\approx 6\,\mu$s across 25 trial medians in 3 sessions) is comparable to within-trial
std, indicating $C_{\mathrm{pi}}$ is not reproducible to that
tolerance across the sessions sampled. As a conservative practice, these results motivate evaluating
per-session recalibration of $C_p$, particularly on Pi, pending
broader replication. The $n=3$
sessions sampled here cannot fully distinguish genuine session drift from
sampling variance; establishing the session-to-session distribution
requires extended replication left to future work.
Jetson's tighter cross-day reproduction ($\sim 0.14\,\mu$s across two sessions) makes
recalibration conservative there; on Pi it is treated as required
pending further data.

The broader implication: cross-platform edge ML timing methodology
may need to be \emph{platform-aware} (separate $C_p$ per platform)
and, depending on observed drift, \emph{session-aware} (recalibrating
when the operating-state envelope is re-established). Gates lacking these properties misclassify
driver-induced overhead as measurement-source inaccuracy, and the
misclassification compounds with the number of platforms compared.

% ---------------------------------------------------------------------------
\section{Limitations and Future Work}
\label{sec:limitations}

This paper is preliminary and limits its claims accordingly.
$C_p$ is characterized only at C0 baseline; behavior under sustained CPU,
memory, I/O, or network contention is left to future work, planned for extension to $n=12$ trials per cell with
multi-condition coverage. If $C_p$ shifts
substantially under contention, Eq.~(\ref{eq:gate}) generalizes to
per-platform-and-per-condition calibration. Statistical replication
here is at the WIP tier ($n=10$ per platform, one session this work);
the Pi cross-day variance is observational, with mechanism (daemon state, thermal
trajectory, etc.) left to future work.

Two GPIO drivers are tested (\texttt{Jetson.GPIO} chardev and
\texttt{pigpio} daemon); other drivers (\texttt{lgpio}, legacy
\texttt{sysfs}) may exhibit different overhead profiles. The single
characterized workload is a $\approx 115$\,k-parameter 1D CNN with
$\approx 1.2$\,ms median inference; sub-millisecond inference times,
where call duration becomes a larger fraction of the per-inference
budget, are out of scope. The Saleae sample rate is fixed at
$100$\,MHz; sample-rate sensitivity of $C_p$ is uncharacterized.

% ---------------------------------------------------------------------------
\section{Conclusion}
\label{sec:conclusion}

Real-time edge ML inference under timing constraints, such as
ECG arrhythmia detection with per-beat budgets, requires that
software-clock latency reflect on-wire timing within validated
tolerances. This work characterized GPIO call overhead in
hardware-validated edge ML inference timing on two embedded Linux
platforms at C0 baseline using such a classifier as the
representative workload. The
per-platform constant differs by approximately $66\,\mu$s between
Jetson and Pi, larger than typical uniform validation tolerances. The
Jetson constant is well-explained by direct call-duration measurement;
the Pi constant is not, motivating empirical per-platform calibration
under the deployed measurement context. Per-platform-constant
validation is a prerequisite for cross-platform edge ML timing
studies. Full validation under contention conditions and at extended
replication tiers is left to future work.

% ---------------------------------------------------------------------------
\bibliography{references}

\end{document}